\documentclass[a4paper]{jpconf}
\usepackage{amssymb}
\usepackage{amsmath}
\usepackage{color}
\usepackage{cite}

\def\blu#1{{\color{blue}#1}}
\def\red#1{{\color{red}#1}}

\def\be{\begin{equation}}
\def\eqn#1{\be\label{#1}}
\def\ee{\end{equation}}

\def\bea{\begin{eqnarray}}
\def\eqnn#1{\bea\label{#1}}
\def\eea{\end{eqnarray}}

\def\md{\medskip}
\def\ta{{\tilde\alpha}}

\def\hh{{\tilde h}}
\def\tk{{\tilde k}}

\def\nn{\nonumber}
\def\nt{\noindent}

\def\np{\vfill\eject}
\def\nd{\end{document}}

\def\tcc{{\tilde{\cal C}}}

\def\han{{\textstyle\frac{n}{2}}}
\def\hel{{\textstyle{11\over2}}}

\input epsf.tex
\newcount\figno
\def\fig#1#2#3{
\par\begingroup\parindent=0pt\leftskip=1cm\rightskip=1cm\parindent=0pt
\baselineskip=11pt \global\advance\figno by 1 
\epsfxsize=#3 \centerline{\epsfbox{#2}} \vskip 12pt
#1\par
\endgroup\par}
\def\figlabel#1{\xdef#1{\the\figno}}
\def\encadremath#1{\vbox{\hrule\hbox{\vrule\kern8pt\vbox{\kern8pt
\hbox{$\displaystyle #1$}\kern8pt} \kern8pt\vrule}\hrule}}

  \def\tV{{\tilde V}}
\def\tcn{{\tilde{\cal N}}}

\def\bu{\noindent $\bullet~$}
\def\rank{{\rm rank}}
\def\deg{{\rm deg}\,}
\def\downcirc#1{\mathop{\circ}\limits_{#1}}
\def\riga{-\kern-4pt - \kern-4pt -}
\font\fat=cmsy10 scaled\magstep5

\def\Bbullet{\raise-3pt\hbox{\fat\char"0F}}

\def\black#1{\mathop{\bullet}\limits_{#1}}

\def\Box{
\vbox{ \halign to5pt{\strut##& \hfil ## \hfil \cr &$\kern -0.5pt
\sqcap$ \cr \noalign{\kern -5pt \hrule} }}~}

\def\down{\raise1.5pt\hbox{$\phantom{a}_2$}\downarrow}

\def\downa{\raise1.5pt\hbox{$\phantom{a}_{2\atop m_2}$}\downarrow}

\def\llr{\longrightarrow}

\def\({\left(}
\def\){\right)}
\def\eps{\epsilon}
 
\def\lra{\longrightarrow}
\def\llra{\longleftrightarrow}

\def\dia{{$\diamondsuit$}}

\def\ha{{\textstyle{\frac{1}{2}}}}

\def\bbz{\mathbb{Z}}
\def\bbc{\mathbb{C}}
\def\bac{\bbc} 
\def\bbr{\mathbb{R}}
\def\bbq{\mathbb{Q}}
\def\bbo{\mathbb{O}}
\def\bbn{\mathbb{N}}
\def\a{\alpha}
\def\b{\beta}

\def\vr{\vert}

\def\ed{\end{document}}

\def\ca{{\cal A}}  \def\cc{{\cal C}}
\def\cd{{\cal D}} \def\ce{{\cal E}} \def\cf{{\cal F}}
\def\cg{{\cal G}} \def\ch{{\cal H}} 
 \def\ck{{\cal K}} 
\def\cm{{\cal M}} \def\cn{{\cal N}} 
\def\cp{{\cal P}}  
 \def\ct{{\cal T}}

\def\ido{intertwining differential operator}
\def\idos{intertwining differential operators}

\def\L{\Lambda}
\def\r{\rho}

\begin{document}

\title{Classification of Invariant Differential Operators for Non-Compact Lie
Algebras via Parabolic Relations\footnote{Talk at  the VIII International Symposium "Quantum Theory and
Symmetries",  Mexico City, August 5-9, 2013.}}

\author{V.K.~Dobrev}

\address{Institute for Nuclear Research and Nuclear Energy,\\
 Bulgarian Academy of Sciences,\\ 72
Tsarigradsko Chaussee, 1784 Sofia, Bulgaria}

\ead{dobrev@inrne.bas.bg}

\begin{abstract}
In the present paper we review the progress of the project of classification and
construction of invariant differential operators for non-compact semisimple Lie groups.
Our starting points is the class of algebras, which we called earlier 'conformal Lie
algebras' (CLA), which have very similar properties to the conformal
algebras of  Minkowski space-time,
though our aim is to go beyond this class in a natural way.  For this
we introduced recently  the new notion of ~{\it parabolic relation}~ between two
non-compact semisimple Lie algebras $\cg$ and $\cg'$ that have the same complexification and
possess maximal parabolic subalgebras with the same complexification. Thus, we consider
 the exceptional algebra   ~$E_{7(7)}$ which is parabolically related to the CLA ~$E_{7(-25)}\,$.
Other interesting examples are the orthogonal algebras ~$so(p,q)$~ all of which are parabolically
related to the conformal  algebra ~$so(n,2)$~ with ~$p+q=n+2$, the parabolic subalgebras including
the Lorentz subalgebra ~$so(n-1,1)$~ and its analogs ~$so(p-1,q-1)$.
Further we consider the algebras $sl(2n,\bbr)$  and for $n=2k$ the algebras $su^*(4k)$
which are parabolically related to the CLA $su(n,n)$.
Further we consider the algebras $sp(r,r)$  which are parabolically related to the CLA $sp(2r,\bbr)$.
We consider also ~$E_{6(6)}$~ and ~$E_{6(2)}$~
which are parabolically related to the hermitian symmetric case ~$E_{6(-14)}\,$.
\end{abstract}

 \section{Introduction}

 \blu{Invariant differential operators} play very important role in
the description of physical symmetries - starting from the early
occurrences in the Maxwell, d'Allembert, Dirac, equations,   to the
latest applications of (super-)differential operators in conformal
field theory, supergravity and string theory (for    reviews, cf. e.g.,
\cite{Mald},\cite{Ter}). Thus, it is important
for the applications in physics to study systematically such
operators. For more relevant references
cf., e.g.,
\cite{Har,BGG,War,Lan,FerrWessZum,Zhea,Kosa,Sokat,FrvNFerr,Wolfa,Ademollo:1976,FayFerr,Wolfb,KnZu,DPPT,OgSok,CrSchFerr,SpVo,Vog,EHW,GIKOS,DobPet,DobN2char,Delduc,TruVar,Jak,KacWak,Kob,Witten,APSW,FHSV,CerDAFerr,AFT,BOO,ABCDFFM,FerrMal,FerrFr,HSW,AGMOO,EdSok,DNW,AEPS,Knab,KacRoWa,FerrSok,Kosb,BaWa,GanVas,Vara,DufFer,FaKoRi,KiMaMi,GuLuMi,HofMal,BCCS,Kallosh,Mizo,FerrKaMa}, and others
throughout the text.

In a recent paper \cite{Dobinv} we started the systematic explicit construction
of invariant differential operators. We gave an explicit description
of the building blocks, namely, the \blu{parabolic subgroups and
subalgebras} from which the necessary representations are induced.
Thus we have set the stage for study of different non-compact
groups.

Since the study and description of detailed classification should be
done group by group we had to decide which groups to study. One
first choice would be non-compact groups that have {\color{blue} discrete series}
of representations. By the Harish-Chandra criterion \cite{HC} these are groups
where holds:
$$ \rank\, G = \rank\ K, $$
where $K$ is the \blu{maximal compact subgroup} of the non-compact group
$G$. Another formulation is to say that the Lie algebra $\cg$ of $G$
has a compact Cartan subalgebra.

\noindent {\it Example:}  the groups \blu{$SO(p,q)$} have discrete series,
\blu{except} when both $p,q$ are \blu{odd} numbers.

This class is rather big, thus, we decided to consider a subclass,
namely, the class of \blu{Hermitian symmetric spaces}. The practical criterion is that in
these cases, the \blu{maximal compact subalgebra} $\ck$ is of the form:
\eqn{hermss} \ck ~=~ so(2) \oplus \ck'   \ee
The Lie algebras from this class are:
\eqn{herm} so(n,2), ~~sp(n,R), ~~su(m,n),  ~~so^*(2n), ~~E_{6(-14)}\,,
~~E_{7(-25)}  \ee These groups/algebras have \blu{highest/lowest weight
representations}, and relatedly {\blu{{\it holomorphic} discrete series
representations}.

The most widely used of these algebras are the \blu{conformal
algebras} ~\red{so(n,2)}~ in $n$-dimensional Minkowski space-time.
In that case, there is a maximal {\blu{Bruhat decomposition} \cite{Bru}:
 that
has direct physical meaning:  \eqnn{bruc} && so(n,2) ~~=~
\blu{\cm\,\oplus\,\ca\,\oplus\,\cn}\, \oplus\,\tcn \ ,\\ && \cm ~=~
so(n-1,1) \ , ~~\dim\ca=1,  ~~\dim \cn = \dim\tcn = n \nn\eea where
~$so(n-1,1)$~ is the \blu{Lorentz algebra} of $n$-dimensional
Minkowski space-time, the subalgebra ~$\ca ~=~ so(1,1)$~ represents
the \blu{dilatations}, the conjugated subalgebras ~$\cn\,$,
$\tcn\,$~ are the algebras of \blu{translations}, and \blu{special
conformal transformations}, both being isomorphic to $n$-dimensional
Minkowski space-time.

The subalgebra ~$\cp = \cm\,\oplus\,\ca\,\oplus\,\cn\,$
 ($\cong \cm\,\oplus\,\ca\,\oplus\,\tcn\,$) is  a \blu{maximal parabolic
 subalgebra}.

There are other special  features which are important. In
particular, the complexification of the maximal compact subgroup is
isomorphic to  the complexification of the first two factors of the
Bruhat decomposition: \eqn{relkm}   \ck^\bac ~=~ {so(n,\bbc)} \oplus
so(2,\bbc) ~\cong~   {so(n-1,1)^\bac} \oplus so(1,1)^\bac = \cm^\bac
\oplus\ca^\bac\ . \ee

In particular, the coincidence of the complexification of the
semi-simple subalgebras:  $$ \red{\ck'^\bac ~=~ \cm^\bac} \eqno(*)$$ means
that the sets of finite-dimensional (nonunitary) representations of
~$\cm$~ are in 1-to-1 correspondence with the finite-dimensional
(unitary) representations of ~$\ck'$. The latter leads to the fact
that the corresponding induced representations
 are representations of finite $\ck$-type \cite{HC}.

It turns out  that some of the hermitian-symmetric algebras  share
the above-mentioned special properties of ~$so(n,2)$.  This subclass
consists of:
\eqn{confc} so(n,2), ~~sp(n,\bbr), ~~su(n,n),  ~~so^*(4n),
~~E_{7(-25)}  \ee   the corresponding analogs of Minkowski
space-time $V$ being:
\eqn{nherm}\bbr^{n-1,1}, ~~{\rm Sym}(n,\bbr), ~~{\rm Herm}(n,\bbc),
~~{\rm Herm}(n,\bbq), ~~ {\rm Herm}(3,\bbo) \ee

 In view of applications to physics, we proposed to call these algebras   '\blu{conformal
Lie algebras}', (or groups).

The corresponding groups are also called '{\it Hermitian symmetric spaces
of tube type}' \cite{FaKo}.
  The same class was identified from different considerations in \cite{Guna}
    called there '{\it conformal groups  of simple Jordan algebras}'.
In fact, the relation between Jordan algebras and division algebras
was known long time ago.   Our class  was identified
 from still different considerations also in \cite{Mackder}    where
they    were called '{\it simple space-time symmetries generalizing
conformal symmetry}'.

We have started the study of the above class in the framework of the
present approach in the cases: ~$so(n,2)$,  ~$su(n,n)$,
~$sp(n,\bbr)$, ~$E_{7(-25)}$,  \cite{Dobpeds}, \cite{Dobsunn}, \cite{Dobspn},  \cite{Dobeseven}, resp.,
and we have considered also the algebra $E_{6(-14)}$, \cite{Dobesix}.

Lately, we discovered an efficient way to extend our considerations
beyond this class introducing the notion of 'parabolically related
non-compact semisimple Lie algebras' \cite{Dobparab}.

\bu {\it Definition:}  ~~~Let ~$\cg,\cg'$~ be two non-compact semisimple Lie algebras
with the same complexification ~$\cg^\bac \cong \cg'^\bac$.
We call them ~\red{parabolically related}~ if they have parabolic subalgebras
~$\cp = \cm \oplus \ca \oplus \cn$,
~$\cp' = \cm' \oplus \ca' \oplus \cn'$,
such that: ~$\cm^\bac ~\cong~ \cm'^\bac$~  ($\Rightarrow \cp^\bac ~\cong~ \cp'^\bac$).\dia

Certainly, there are many such parabolic relationships for any given algebra ~$\cg$.
Furthermore, two algebras ~$\cg,\cg'$~ may be parabolically related via different
parabolic subalgebras.

We summarize the algebras parabolically related to conformal Lie
algebras with maximal parabolics fulfilling $(*)$  in the following
table:

\bigskip

\fig{}{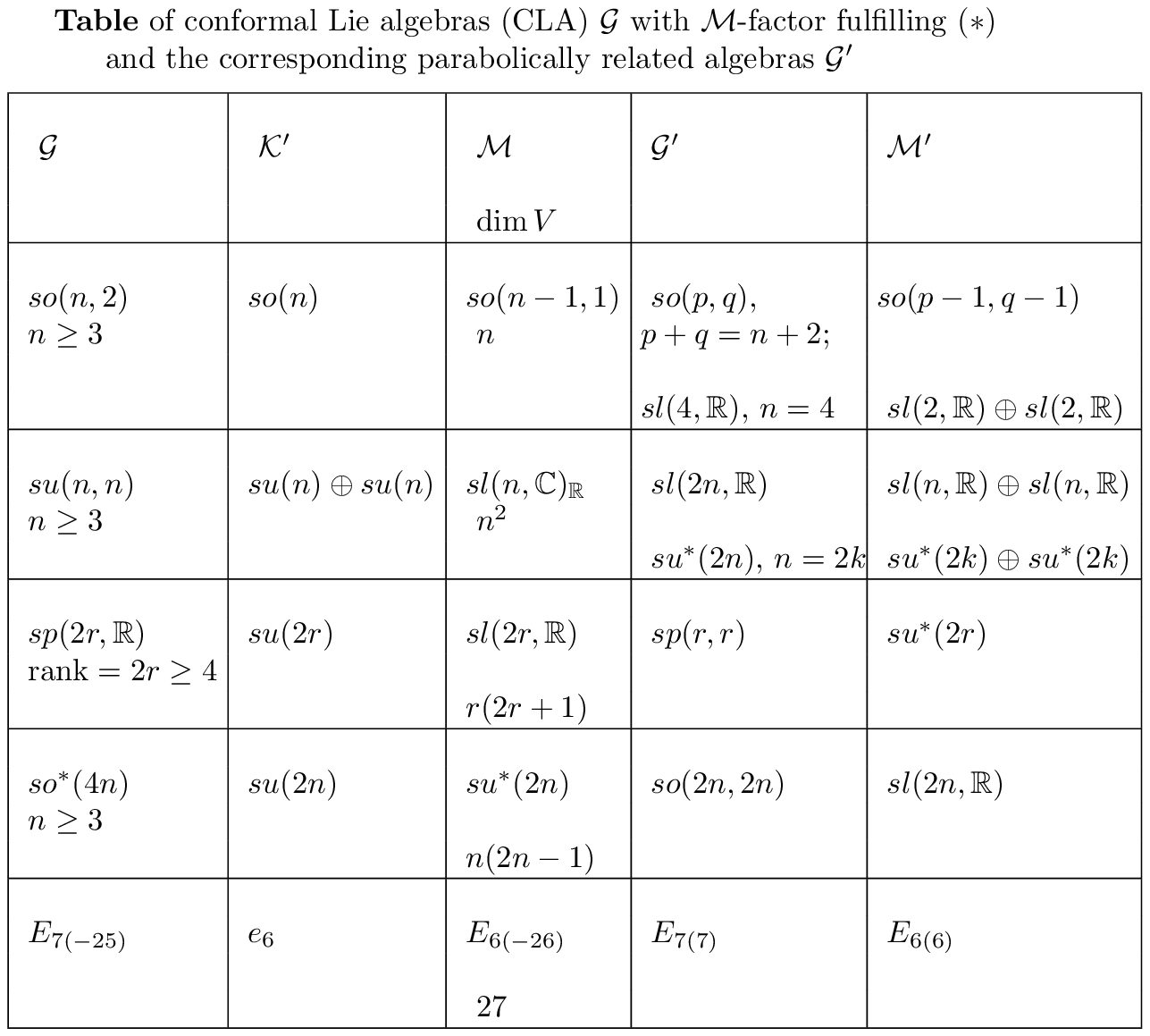}{18cm}
\nt where we display only the semisimple
part ~$\ck'$~ of ~$\ck$; ~$sl(n,\bbc)_\bbr$~ denotes $sl(n,\bbc)$ as
a real Lie algebra, (thus, ~$(sl(n,\bbc)_\bbr)^\bac =
sl(n,\bbc)\oplus sl(n,\bbc)$); ~$e_6$~ denotes the compact real form
of ~$E_6\,$; and we have imposed
restrictions to avoid coincidences or degeneracies due to well known isomorphisms:
~$so(1,2) \cong sp(1,\bbr) \cong su(1,1) $, ~~$so(2,2) \cong so(1,2)\oplus so(1,2)$,
~$su(2,2) \cong so(4,2)$, ~$sp(2,\bbr) \cong so(3,2)$,
  ~$so^*(4)\cong so(3) \oplus so(2,1)$, ~$so^*(8)\cong so(6,2)$.

\md

After this extended introduction we give the outline of the paper.
In Section 2 we give the
preliminaries, actually recalling and adapting facts from
\cite{Dobinv}.
In Section 3 we consider   the case of the pseudo-orthogonal algebras
$so(p,q)$ which are parabolically related to the conformal algebra $so(n,2)$
for $p+q=n+2$.
In Section 4 we consider the CLA  ~$su(n,n)$~ and the parabolically
related  $sl(2n,\bbr)$, and for ~$n=2k$~:~ $su^*(4k)$.
In Section 5 we consider the CLA  ~$sp(n)$~ and - for  ~$n=2r$~ - the parabolically related $sp(r,r)$.
  In Section 6 we consider the CLA  ~$E_{7(-25)}$~ and the parabolically
related  $E_{7(7)}$.
In Section 7 we consider the hermitian symmetric case ~$E_{6(-14)}\,$ and  the parabolically
related   $E_{6(6)}$ and $E_{6(2)}$.

\section{Preliminaries}

 Let $G$ be a semisimple non-compact Lie group, and $K$ a
maximal compact subgroup of $G$. Then we have an {\it Iwasawa
decomposition} ~$G=KA_0N_0$, where ~$A_0$~ is Abelian simply
connected vector subgroup of ~$G$, ~$N_0$~ is a nilpotent simply
connected subgroup of ~$G$~ preserved by the action of ~$A_0$.
Further, let $M_0$ be the centralizer of $A_0$ in $K$. Then the
subgroup ~$P_0 ~=~ M_0 A_0 N_0$~ is a {\it minimal parabolic subgroup} of
$G$.  A {\it parabolic subgroup} ~$P ~=~ M' A' N'$~ is any subgroup of $G$
which contains a minimal parabolic subgroup.

Further, let ~$\cg,\ck,\cp,\cm,\ca,\cn$~ denote the Lie algebras of ~$G,K,P,M,A,N$, resp.

For our purposes we need to restrict to  ~{\it maximal ~ parabolic
subgroups} ~$P=MAN$, i.e.  $\rank A =1$, resp. to ~{\it maximal ~ parabolic
subalgebras} ~$\cp = \cm \oplus \ca \oplus \cn$~ with ~$\dim\, \ca=1$.

Let ~$\nu$~ be a (non-unitary) character of ~$A$, ~$\nu\in\ca^*$,
parameterized by a real number ~{\it $d$}, called the {\it conformal weight} or
energy.

Further, let ~ $\mu$ ~ fix a discrete series representation
~$D^\mu$~ of $M$ on the Hilbert space ~$V_\mu\,$, or   the
finite-dimensional (non-unitary) representation of $M$ with the same
Casimirs.

 We call the induced
representation ~$\chi =$ Ind$^G_{P}(\mu\otimes\nu \otimes 1)$~ an
~\blu{\it elementary representation} of $G$ \cite{DMPPT}. (These are
called {\it generalized principal series representations} (or {\it
limits thereof}) in \cite{Knapp}.)   Their spaces of functions are:  \eqn{func}
\cc_\chi ~=~ \{ \cf \in C^\infty(G,V_\mu) ~ \vr ~ \cf (gman) ~=~
e^{-\nu(H)} \cdot D^\mu(m^{-1})\, \cf (g) \} \ee where ~$a=
\exp(H)\in A'$, ~$H\in\ca'\,$, ~$m\in M'$, ~$n\in N'$. The
representation action is the \blu{left regular action}:  \eqn{lrega}
(\ct^\chi(g)\cf) (g') ~=~ \cf (g^{-1}g') ~, \quad g,g'\in G\ .\ee

\bu An important ingredient in our considerations are the ~\blu{\it
highest/lowest weight representations}~ of ~$\cg^\bac$. These can be
realized as (factor-modules of) Verma modules ~$V^\L$~ over
~$\cg^\bac$, where ~$\L\in (\ch^\bac)^*$, ~$\ch^\bac$ is a Cartan
subalgebra of ~$\cg^\bac$, weight ~$\L = \L(\chi)$~ is determined
uniquely from $\chi$ \cite{Dob}.

Actually, since our ERs may be induced from finite-dimensional
representations of ~$\cm$~ (or their limits) the Verma modules are
always reducible. Thus, it is more convenient to use ~\blu{\it
generalized Verma modules} ~$\tV^\L$~ such that the role of the
highest/lowest weight vector $v_0$ is taken by the
(finite-dimensional) space ~$V_\mu\,v_0\,$. For the generalized
Verma modules (GVMs) the reducibility is controlled only by the
value of the conformal weight $d$. Relatedly, for the \idos{} only
the reducibility w.r.t. non-compact roots is essential.

\bu Another main ingredient of our approach is as follows. We group
the (reducible) ERs with the same Casimirs in sets called \red{~{\it
multiplets}} \cite{Dob}. The multiplet corresponding to fixed values of the
Casimirs may be depicted as a connected graph, the \blu{vertices} of
which correspond to the reducible ERs and the \blu{lines (arrows)}
between the vertices correspond to intertwining operators.  The
explicit parametrization of the multiplets and of their ERs is
important for understanding of the situation. The notion of multiplets was introduced in \cite{Dobmul},\cite{Dobc}
and applied to representations of ~$SO_o(p,q)$~ and ~$SU(2,2)$, resp.,
induced from their minimal parabolic subalgebras. Then it was applied to
the conformal superalgebra \cite{DoPemul}, to
infinite-dimensional (super-)algebras \cite{Dobmuinf},
to quantum groups \cite{Dobqg}. (For other applications
we refer to \cite{Dobmuvar}.)

In fact, the multiplets contain explicitly all the data necessary to
construct the \idos{}. Actually, the data for each \ido{} consists
of the pair ~$(\b,m)$, where $\b$ is a (non-compact) positive root
of ~$\cg^\bac$, ~$m\in\bbn$, such that the \blu{BGG  Verma module
reducibility condition} (for highest weight modules) is fulfilled:
\eqn{bggr} (\L+\r, \b^\vee ) ~=~ m \ , \quad \b^\vee \equiv 2 \b
/(\b,\b) \ \ee
$\r$ is half the sum of the positive roots of
~$\cg^\bac$. When the above holds then the Verma module with shifted
weight ~$V^{\L-m\b}$ (or ~$\tV^{\L-m\b}$ ~ for GVM and $\b$
non-compact) is embedded in the Verma module ~$V^{\L}$ (or
~$\tV^{\L}$). This embedding is realized by a singular vector
~$v_s$~ determined by a polynomial ~$\cp_{m,\b}(\cg^-)$~ in the
universal enveloping algebra ~$(U(\cg_-))\ v_0\,$, ~$\cg^-$~ is the
subalgebra of ~$\cg^\bac$ generated by the negative root generators
\cite{Dix}.
 More explicitly, \cite{Dob}, ~$v^s_{m,\b} = \cp_{m,\b}\, v_0$ (or ~$v^s_{m,\b} = \cp_{m,\b}\, V_\mu\,v_0$ for GVMs).
   Then
there exists \cite{Dob} an \red{\ido{}} \eqn{invop}  \cd_{m,\b} ~:~ \cc_{\chi(\L)}
~\llr ~ \cc_{\chi(\L-m\b)} \ee given explicitly by: \eqn{singvv}
 \cd_{m,\b} ~=~ \cp_{m,\b}(\widehat{\cg^-})  \ee where
~$\widehat{\cg^-}$~ denotes the \blu{right action} on the functions
~$\cf$.

In most of these situations the invariant operator ~$\cd_{m,\b}$~ has a non-trivial invariant
kernel in which a subrepresentation of $\cg$ is realized. Thus, studying the equations
with trivial RHS:
\eqn{invdec} \cd_{m,\b}\ f ~=~ 0 \ , \qquad f \in \cc_{\chi(\L)} \ ,\ee
is also very important. For example, in many physical applications
 in the case of first order differential operators,
i.e., for ~$m=m_\b = 1$, these equations
are called ~\blu{conservation laws}, and the elements ~$f\in \ker \cd_{m,\b}$~
are called ~\blu{conserved currents}.

The above construction works also for the ~\blu{subsingular
vectors}~ $v_{ssv}$~ of Verma modules. Such a vector is also
expressed by a polynomial ~$\cp_{ssv}(\cg^-)$~ in the universal
enveloping algebra:
   ~$v^s_{ssv} = \cp_{ssv}(\cg^-)\, v_0\,$, cf. \cite{Dobcond}.
Thus, there exists a ~{\it conditionally invariant differential operator}
~ given explicitly by:
~$\cd_{ssv} ~=~ \cp_{ssv}(\widehat{\cg^-})$,
and a ~{\it conditionally invariant differential equation},
for many more details, see \cite{Dobcond}.
(Note that these  operators (equations) are not of first order.)

Below in our exposition we shall use the so-called Dynkin labels: \eqn{dynk} m_i
~\equiv~ (\L+\r,\a^\vee_i)  \ , \quad i=1,\ldots,n, \ee where ~$\L =
\L(\chi)$, ~$\r$ is half the sum of the positive roots of
~$\cg^\bac$.

We shall use also   the so-called Harish-Chandra parameters:
\eqn{dynhc} m_\b \equiv (\L+\r, \b )\ ,  \ee where $\b$ is any
positive root of $\cg^\bac$. These parameters are redundant, since
they are expressed in terms of the Dynkin labels, however,   some
statements are best formulated in their terms. (Clearly, both the Dynkin labels and
Harish-Chandra parameters have their origin in the BGG reducibility condition \eqref{bggr}.)

\section{Conformal algebras  ~\blu{$so(n,2)$}~ and parabolically
related}

\nt Let ~$\cg=so(n,2)$,  $n>2$. We label   the signature of the ERs
of $\cg$   as follows: \eqnn{sgnd}  &&\chi ~~=~~ \{\, n_1\,,
\ldots,\, n_{\hh}\,;\, c\, \} \ , \quad n_j \in \bbz/2\ , \quad
c=d-\han\ , \quad \hh \equiv [\han] ,\\ && \vr n_1 \vr < n_2 <
\cdots <  n_{\hh}\ , \quad n ~{\rm even}\ ,\nn\\ && 0 < n_1 < n_2 <
\cdots <  n_{\hh} \ , \quad n ~{\rm odd}\ ,\nn\eea where the last
entry of ~$\chi$~ labels the characters of $\ca\,$, and the first
$\hh$ entries are labels of the finite-dimensional nonunitary irreps
of $\cm\cong so(n-1,1) $.

The reason to use the parameter ~$c$~ instead of ~$d$~ is that the
parametrization of the ERs in the multiplets is given in a simple
 intuitive way (cf. \cite{Dobsrni},\cite{Dobpeds}):
\eqnn{sgne}  \chi^\pm_1 &=& \{ \eps  n_1\,, \ldots,\,
n_\hh \,;\, \pm n_{\hh+1} \} \ ,   \quad n_\hh < n_{\hh+1}\ , \\
\chi^\pm_2 &=& \{ \eps  n_1\,, \ldots,\, n_{\hh-1}\,,\,
n_{\hh+1}\,;\, \pm n_\hh \}     \nn\\  \chi^\pm_3 \!&=&\! \{\! \eps
n_1,\! \ldots,\! n_{\hh-2},\! n_{\hh},\! n_{\hh+1}\,;\, \pm
n_{\hh-1} \}     \nn\\  ... \nn\\  \chi^\pm_{\hh} &=& \{ \eps n_1\,,
n_3\,, \ldots,\, n_{\hh}\,,\, n_{\hh+1}\,;\, \pm  n_2 \}     \nn\\
\chi^\pm_{\hh+1} &=& \{ \eps n_2\,, \ldots,\, n_{\hh}\,,\,
n_{\hh+1}\,;\, \pm  n_1 \}     \nn\\  &\eps =& \begin{cases}
 \pm\,, ~&~
  n ~ even  \nn\\
                     1,  ~&~  n  ~ odd \end{cases} \nn\eea

 Further, we denote by ~$\tcc^\pm_i$~ the representation space with signature
~$\chi^\pm_i\,$.

The number of ERs in the corresponding multiplets is
equal to:
\eqn{multi} \vr W(\cg^\bac,\ch^\bac)\vr\, /\, \vr
W(\cm^\bac,\ch_m^\bac)\vr    ~=~ 2(1+\hh) \ee
where ~$\ch^\bac,\ \ch^\bac_m$~ are Cartan subalgebras of ~$\cg^\bac,\ \cm^\bac$, resp.
This formula is valid for the main multiplets of all conformal Lie algebras.

We show some examples of diagrams of invariant differential operators
for the conformal groups $so(5,1)$, resp. $so(4,2)$,  in 4-dimensional Euclidean, resp. Minkowski, space-time.
 In Fig.~1. we show the simplest example for the most common using well known operators.
 In Fig.~2. we show the same example but using the group-theoretical parity splitting of the electromagnetic current,
 cf. \cite{DoPe:78}.
In Fig.~3. we show the general classification for ~$so(5,1)$ given in \cite{DoPe:78}.
These diagrams are valid also for $so(4,2)$ \cite{PeSo} and for ~$so(3,3)\cong sl(4,\bbr)$ \cite{Dobparab}.

\bigskip

\fig{}{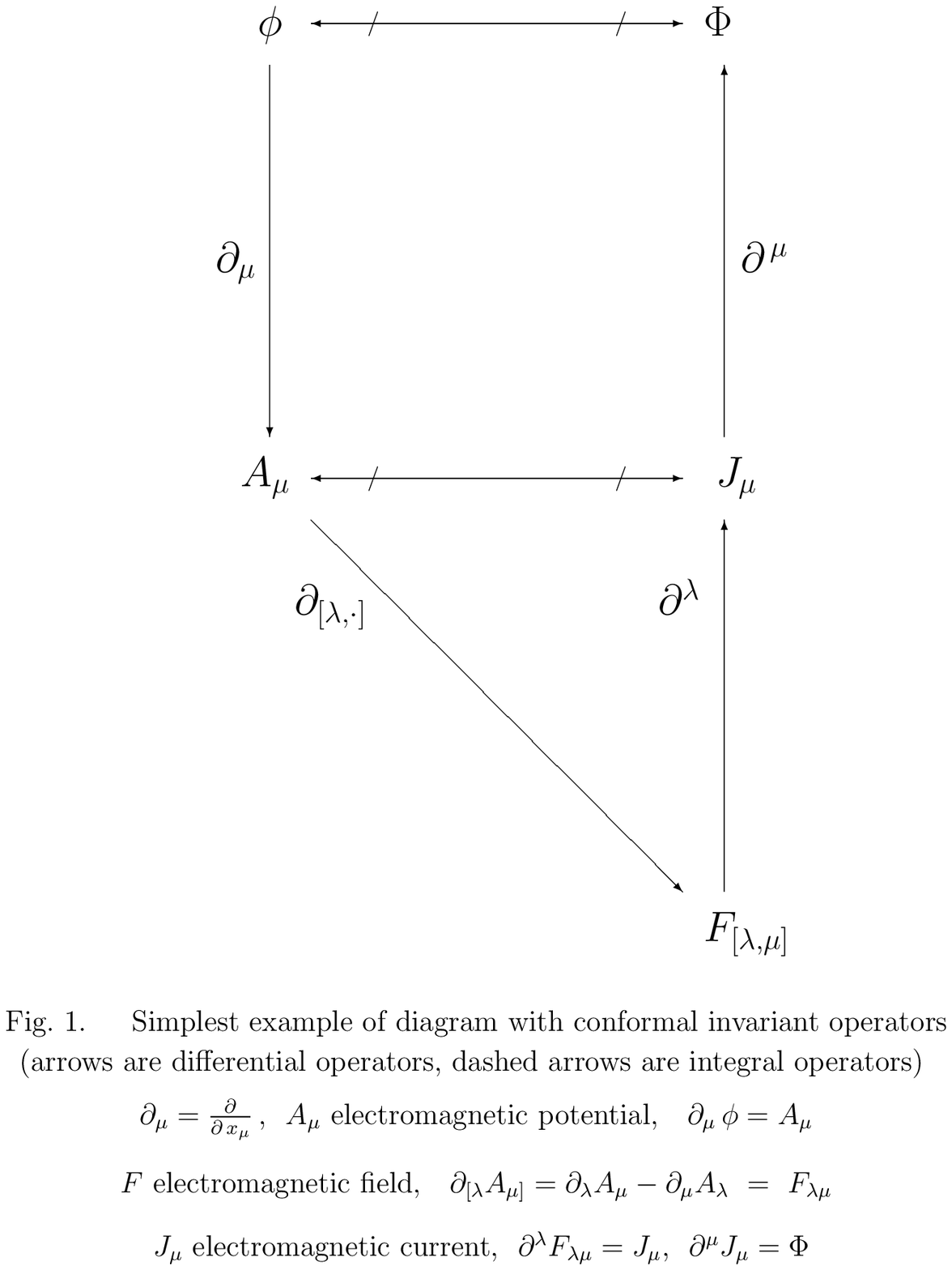}{10cm}

\fig{}{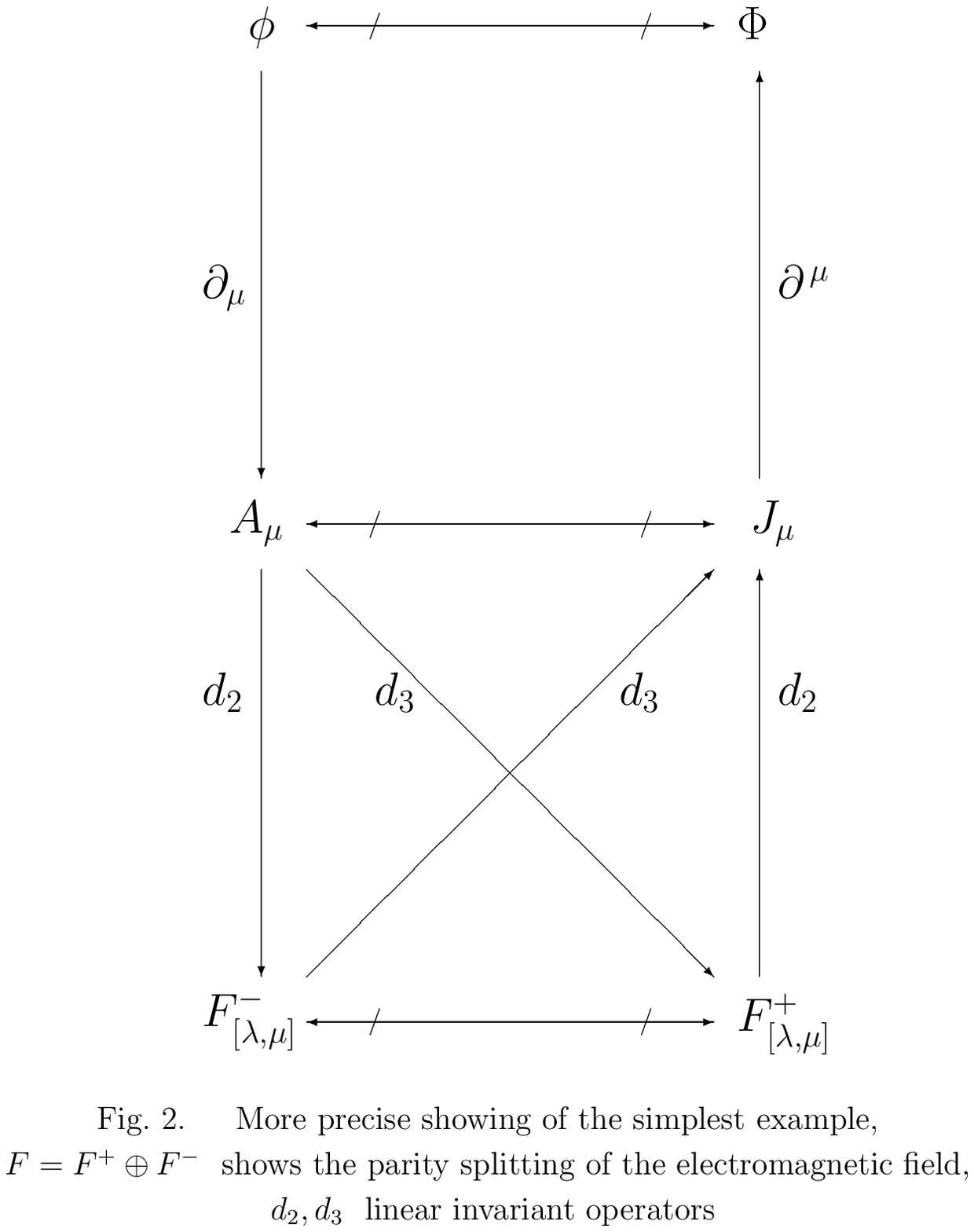}{8cm}

\fig{}{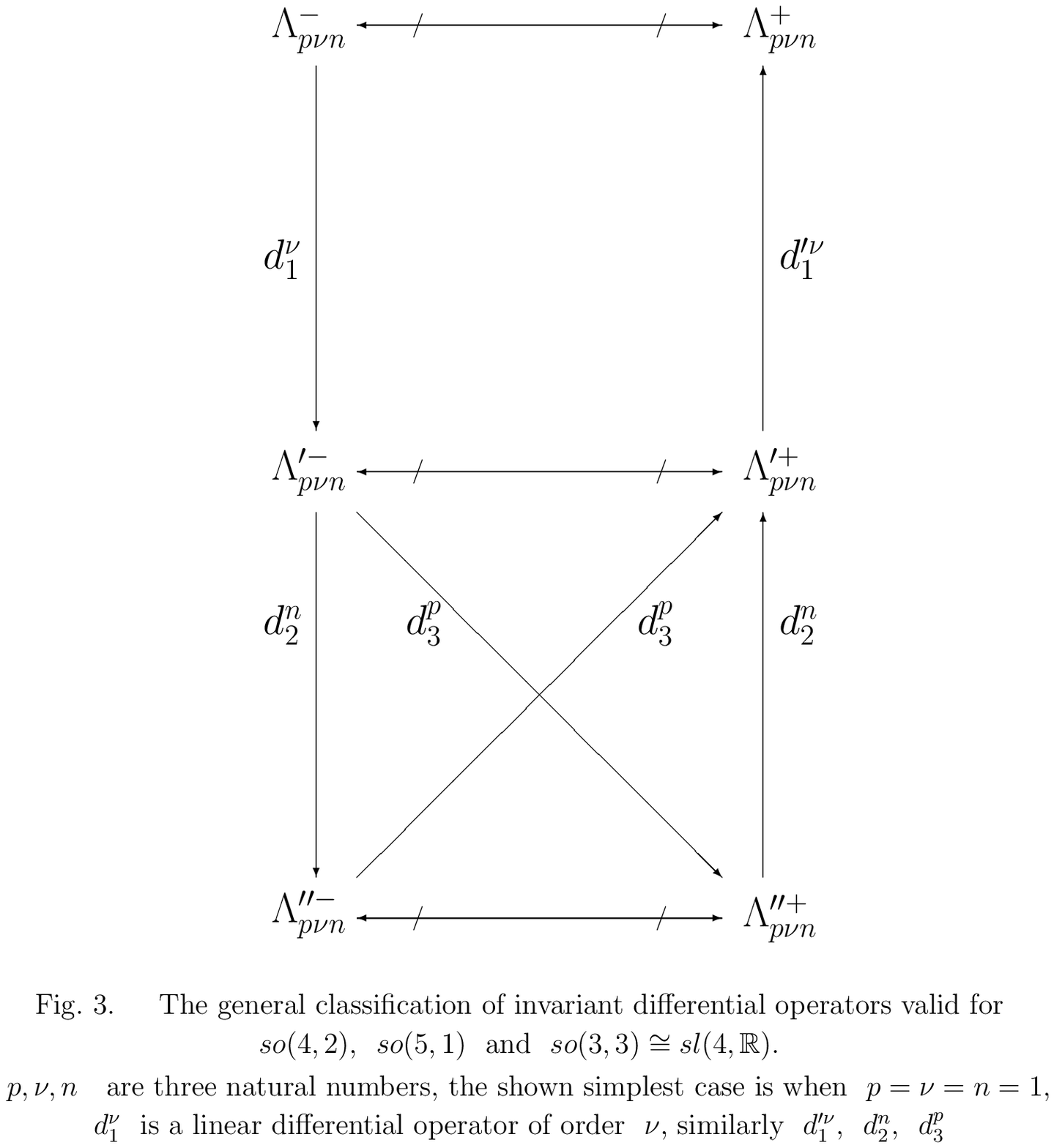}{9cm}

\np

Next in Fig.~4. we show the general even case~   $so(p,q)$, $p+q=2h+2$-even, \cite{Dobsrni},\cite{Dobpeds}, while in Fig.~5.
we show an alternative view of the same case:

\bigskip

\fig{}{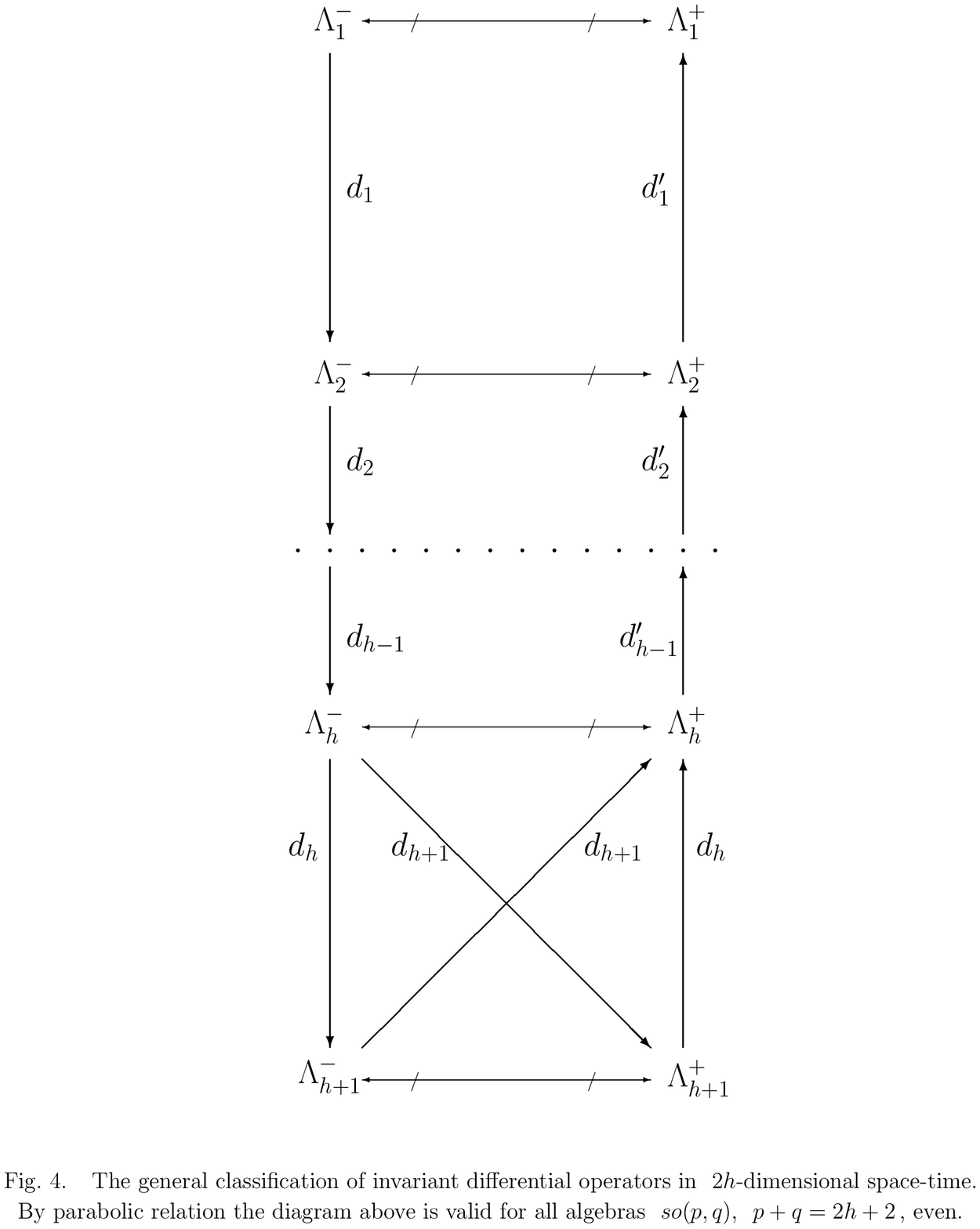}{9cm}

\fig{}{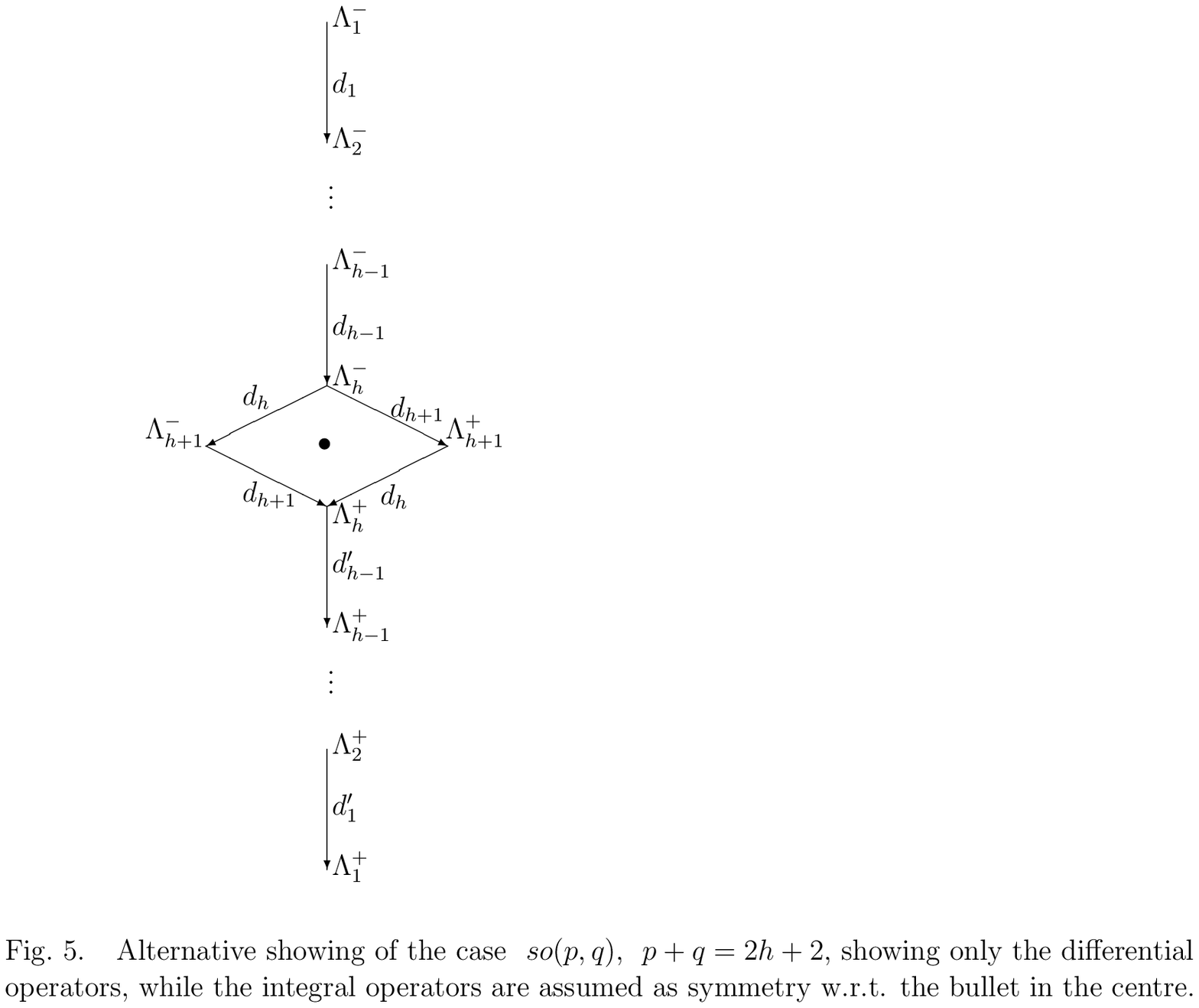}{11cm}

\np

Next in Fig.~6. we show the general odd case~   $so(p,q)$, $p+q=2h+3$-odd, \cite{Dobsrni},\cite{Dobpeds}, while in Fig.~7.
we show an alternative view of the same case:

\bigskip

\fig{}{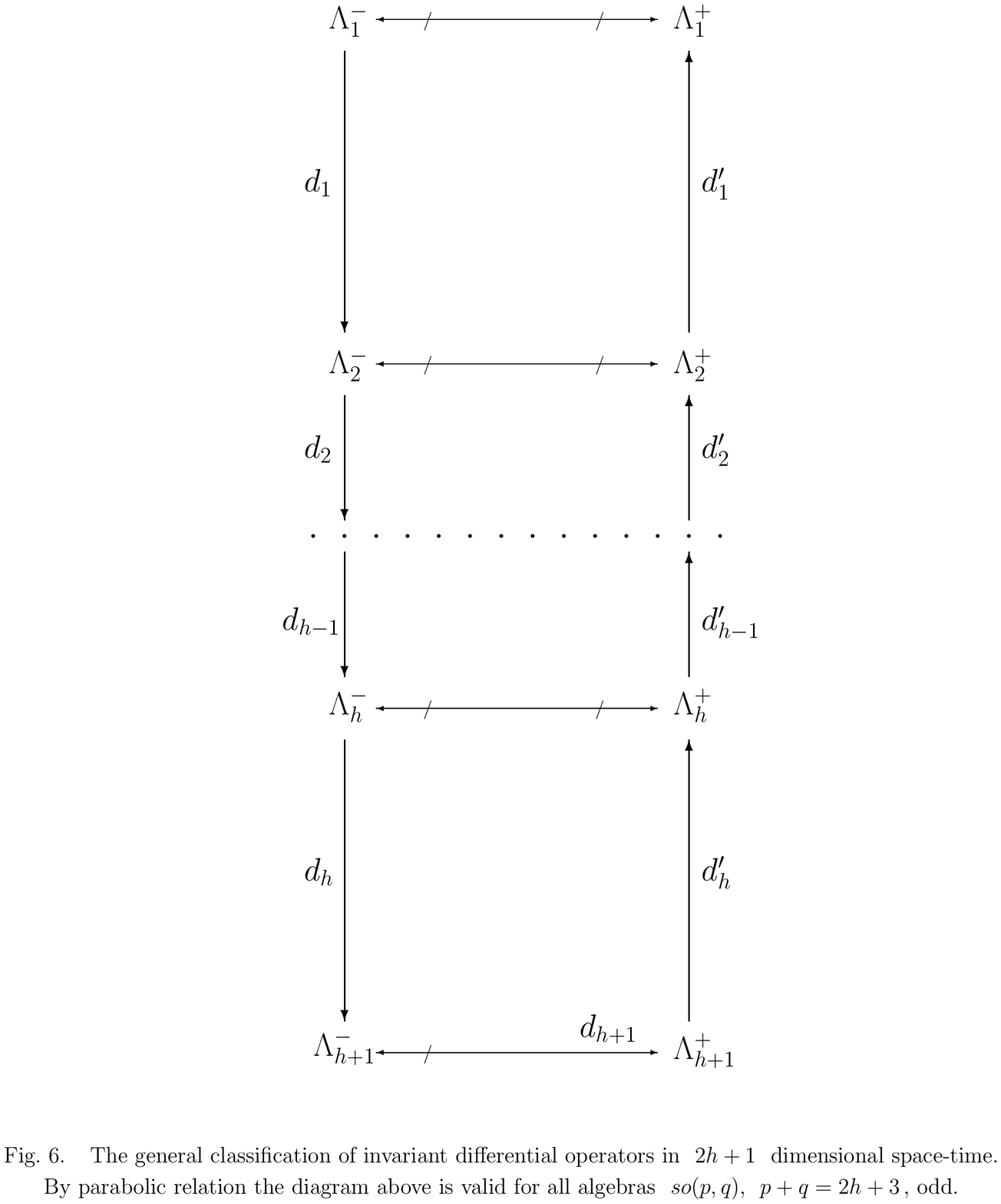}{9cm}

\fig{}{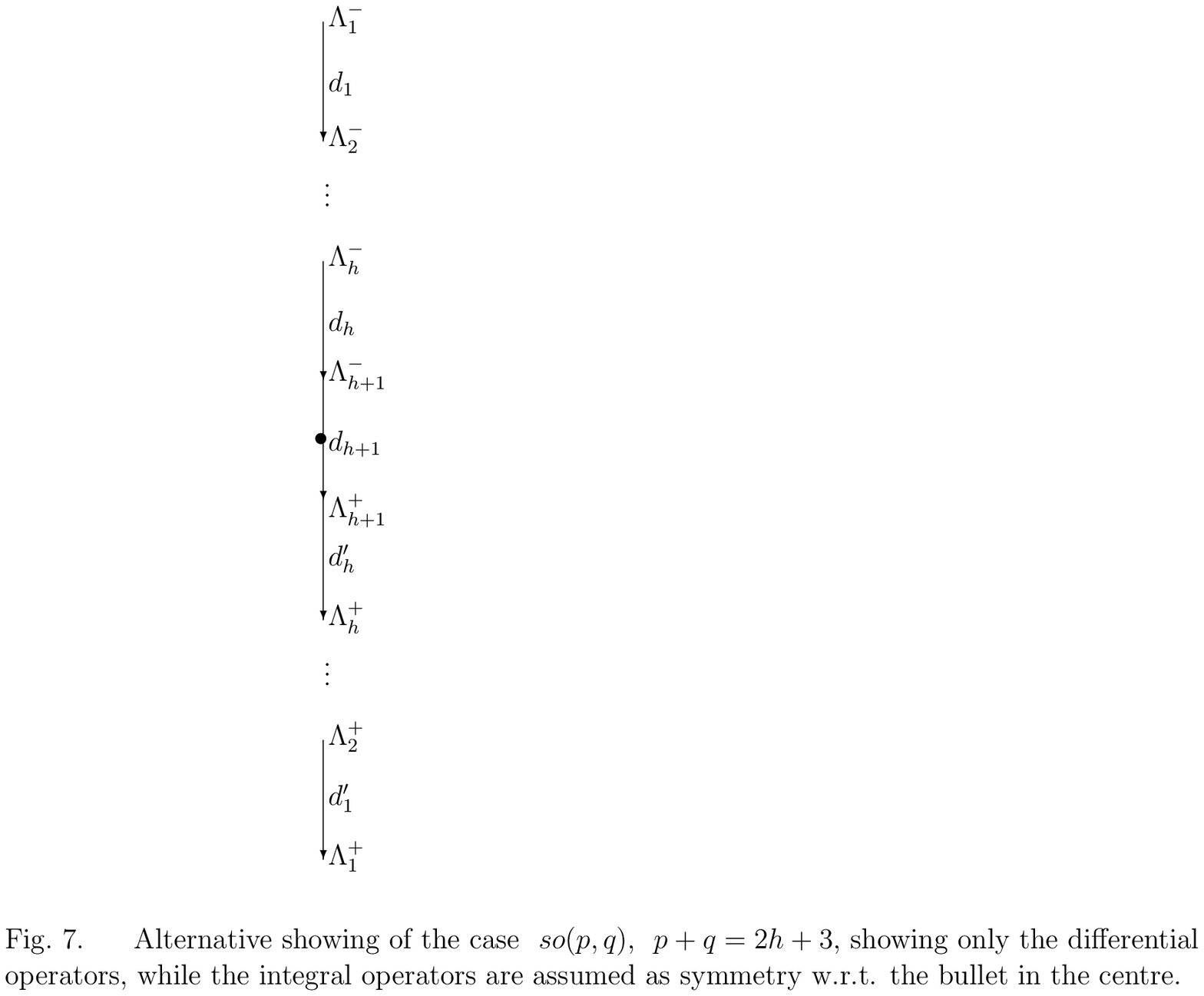}{11cm}

\np

The ERs in the multiplet are related by \blu{intertwining integral and
differential operators}. The  \blu{integral operators} were introduced by
Knapp and Stein \cite{KnSt}. They correspond to elements of the restricted Weyl group of $\cg$.
These operators intertwine the pairs ~$\tcc^\pm_i$~
\eqn{knapps}
G^\pm_i ~:~ \tcc^\mp_i \lra \tcc^\pm_{i}   \ , \quad i ~=~
1,\ldots,1+\hh  \ee

The \blu{\idos}\ correspond to non-com\-pact positive roots of the root
system of ~$so(n+2,\bbc)$, cf. \cite{Dob}. [In the current context, compact
roots of $so(n+2,\bbc)$ are those that are roots also of the
subalgebra $so(n,\bbc)$, the rest of the roots are non-compact.]
The degrees of these
\idos\ are given just by the differences of the ~$c$~ entries
\cite{Dobsrni}: \eqnn{degr} &&\deg d_i  =  \deg d'_i = n_{\hh+2-i} -
n_{\hh+1-i} \,,  \qquad i = 1,\ldots,\hh \,, \quad \forall n \\
&&\deg d_{\hh+1} = n_2 + n_1   \,, \quad n ~ {\rm even} \nn\eea
where $d'_h$ is omitted from the first line for $(p+q)$ even.

Matters are arranged so that in every multiplet only the ER with
signature ~$\chi^-_1$~ contains a \blu{finite-dimensional nonunitary
subrepresentation} in  a   subspace ~$\ce$. The
latter corresponds to the finite-dimensional unitary irrep of
~$so(n+2)$~ with signature ~$\{ n_1\,, \ldots,\, n_\hh \,, \,
n_{\hh+1} \}$. The subspace ~$\ce$~ is annihilated by the operator
~$G^+_1\,$,\ and is the image of the operator ~$G^-_1\,$.

Although the diagrams are valid for arbitrary  $so(p,q)$ ($p+q\geq 5$) the contents
is very different. We comment only on the ER with signature
~$\chi^+_1\,$. In all cases it contains  an UIR of $so(p,q)$ realized on an
invariant subspace ~$\cd$~ of
the ER ~$\chi^+_1\,$. That subspace is annihilated by the operator
~$G^-_1\,$,\ and is the image of the operator ~$G^+_1\,$.
(Other ERs contain more UIRs.)

If ~$pq \in 2\bbn$~ the mentioned UIR is a discrete series representation.
(Other ERs contain more discrete series UIRs.)

And if ~$q=2$~ the invariant subspace ~$\cd$~ is the direct sum of two subspaces
~$\cd ~=~ \cd^+ \oplus \cd^-$, in which are realized a
 {\it holomorphic discrete series representation} and its conjugate
  {\it   anti-holomorphic discrete
series representation}, resp.
Note that the corresponding \blu{lowest weight GVM} is infinitesimally
equivalent only to the holomorphic discrete series, while the
conjugate \blu{highest weight GVM} is infinitesimally equivalent to the
anti-holomorphic discrete series.

 Note that
the  ~$\deg d_i\,$, ~$\deg d'_i\,$, are
Harish-Chandra parameters   corresponding to the non-compact
positive roots of ~$so(n+2,\bbc)$. From these, only  $\deg d_1\,$
corresponds to a simple root, i.e., is a Dynkin label.

Above we considered ~$so(n,2)$~ for  ~$n>2$. The case  ~$n=2$~ is reduced to
~$n=1$~ since ~$so(2,2) \cong so(1,2) \oplus so(1,2)$. The case
\blu{~$so(1,2)$~} is special   and must be  treated separately. But in fact,  it is
contained in what we presented already. In that case the multiplets contain only \blu{two ERs} which
may be depicted by the \blu{top pair $\chi^\pm_1$} in the pictures that we presented. And they have the properties
that we described for ~$so(n,2)$~ with $n>2$.
The case ~$so(1,2)$~ was given already in 1946-7 independently by
Gel'fand et al \cite{GeNa} and Bargmann \cite{Barg}.

\section{The  Lie algebra \blu{$su(n,n)$} and parabolically
related}

\nt Let ~$\cg ~=~ su(n,n)$, ~$n\geq 2$. The maximal compact subgroup
is ~$\ck \cong u(1)\oplus su(n)\oplus su(n)$, while ~$\cm =
sl(n,\bbc)_\bbr\,$.  The number of ERs in the corresponding
multiplets is equal to $$\vr W(\cg^\bac,\ch^\bac)\vr\, /\, \vr
W(\cm^\bac,\ch_m^\bac)\vr ~=~ \left( {2n\atop n}\right)$$ The
signature of the ERs of $\cg$    is: \eqnn{}   \chi ~&=&~ \{\,
n_1\,, \ldots,\, n_{n-1}\,,\, n_{n+1}\, \ldots,\, n_{2n-1}\,;\, c\,
\} \ ,   \quad n_j \in \bbn\ , \quad c = d- n \nn\eea  The
Knapp--Stein   restricted Weyl reflection is given by:
 \eqnn{knast}  && G_{KS} ~:~ \cc_\chi ~ \llr ~ \cc_{\chi'} \
, \quad \chi'  = \{
(n_1,\ldots,n_{n-1},n_{n+1},\ldots,n_{2n-1})^*  ;  -c  \} \qquad\nn \\
&&  (n_1,\ldots,n_{n-1},n_{n+1},\ldots,n_{2n-1})^* ~\doteq~
 (n_{n+1},\ldots,n_{2n-1},n_1,\ldots,n_{n-1}) \nn\eea

Below  in Fig.~8 and  in Fig.~9  we give the diagrams for ~$su(n,n)$  for ~$n=3,4$, \cite{Dobsunn}.
(The case $n=2$ is already considered since $su(2,2)\cong so(4,2)$.)
These are diagrams also for the parabolically related
~$sl(2n,\bbr)$, and for ~$n=2k$~ these are diagrams also for the parabolically related
~$su^*(4k)$, \cite{Dobparab}.

We use the following conventions. Each \ido\ is represented by an
arrow accompanied by a symbol ~$i_{j...k}$~ encoding the root
~$\b_{j...k}$~ and the number $m_{\b_{j...k}}$ which is involved in
the BGG criterion.

\bigskip

\fig{}{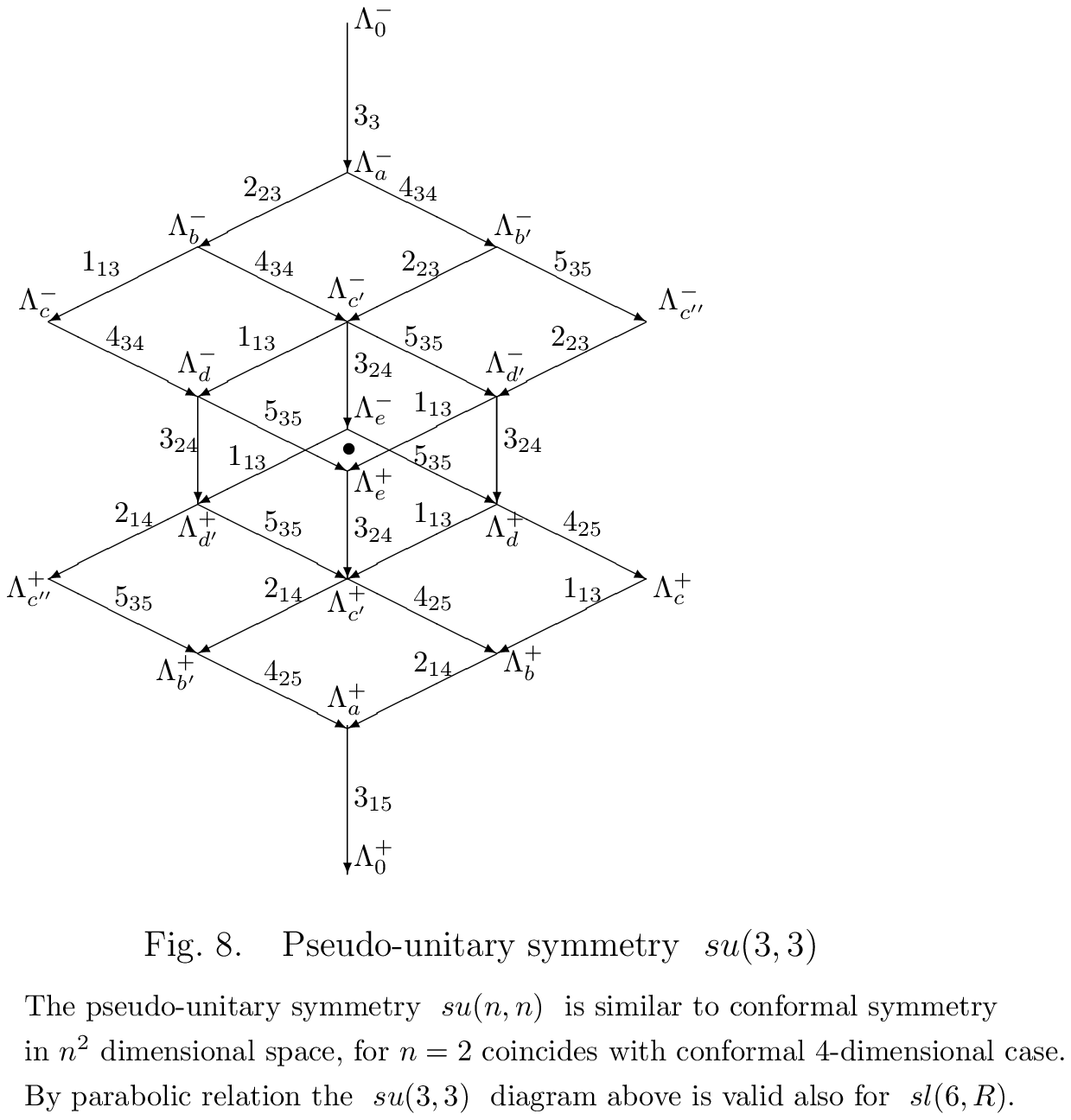}{14cm}

\fig{}{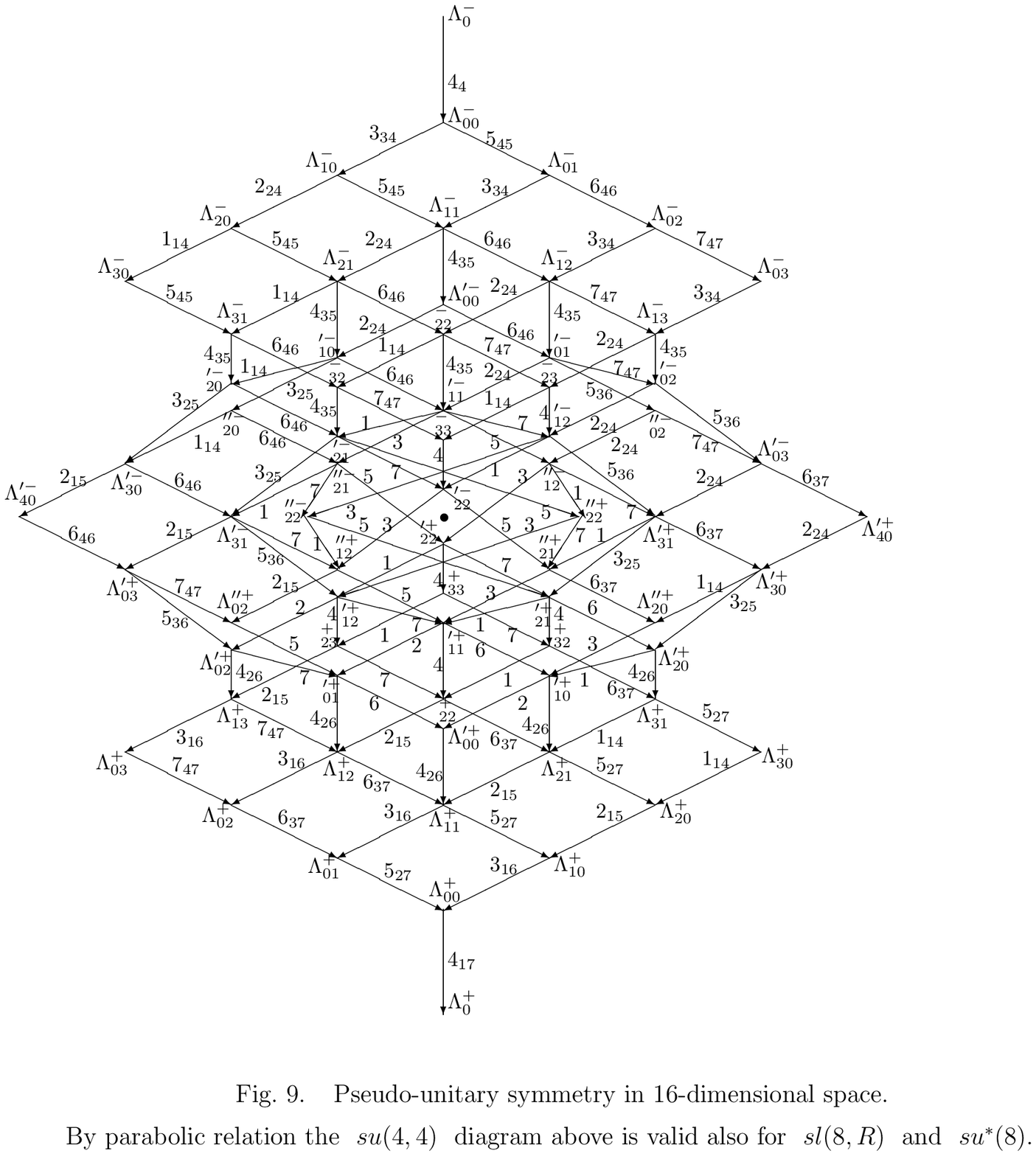}{20cm}

\section{The  Lie algebras \blu{$sp(n,\bbr)$} and
\blu{$sp(\han,\han)$}  ($n$--even)}

\nt Let ~$n\geq 2$. Let ~$\cg ~=~ sp(n,\bbr)$, the split real form
of ~$sp(n,\bbc)=\cg^\bac$. The maximal compact subgroup is ~$\ck
\cong u(1)\oplus su(n)$, while ~$\cm = ~sl(n,\bbr)$.
 The number of ERs in the corresponding multiplets is:   $$\vr
W(\cg^\bac,\ch^\bac)\vr\, /\, \vr W(\cm^\bac,\ch_m^\bac)\vr ~=~ 2^n$$
 The signature of the ERs of $\cg$   is:
 $$  \chi ~=~ \{\, n_1\,, \ldots,\, n_{n-1}\, ;\, c\, \} \ ,
\qquad n_j \in \bbn\ , $$
The Knapp-Stein    Weyl reflection  acts as follows:
  \eqnn{} && G_{KS} ~:~ \cc_\chi ~ \llr ~ \cc_{\chi'} \ ,
  \chi' ~=~
\{\, (n_1,\ldots,n_{n-1})^* \,;\, -c\, \} \ ,  \nn \\
   &&(n_1,\ldots,n_{n-1})^* ~\doteq~ (n_{n-1},\ldots,n_{1})  \nn \eea

Below in Fig.~10, Fig.~11, Fig.~12  and  Fig.~13
we give pictorially the multiplets for ~$sp(n,\bbr)$~ for ~$n=3,4,5,6$, \cite{Dobspn}.
(The case $n=2$ is already considered since $sp(2,\bbr)\cong so(3,2)$.)
For ~$n=2r$~   these are also  multiplets for ~$sp(r,r)$, ~$r=1,2,3$, \cite{Dobparab}.
(The case $n=2,r=1$ is already considered due to $sp(1,1)\cong so(4,1)$ and the parabolic relation
between $so(3,2)$ and $so(4,1)$.)

\bigskip

\fig{}{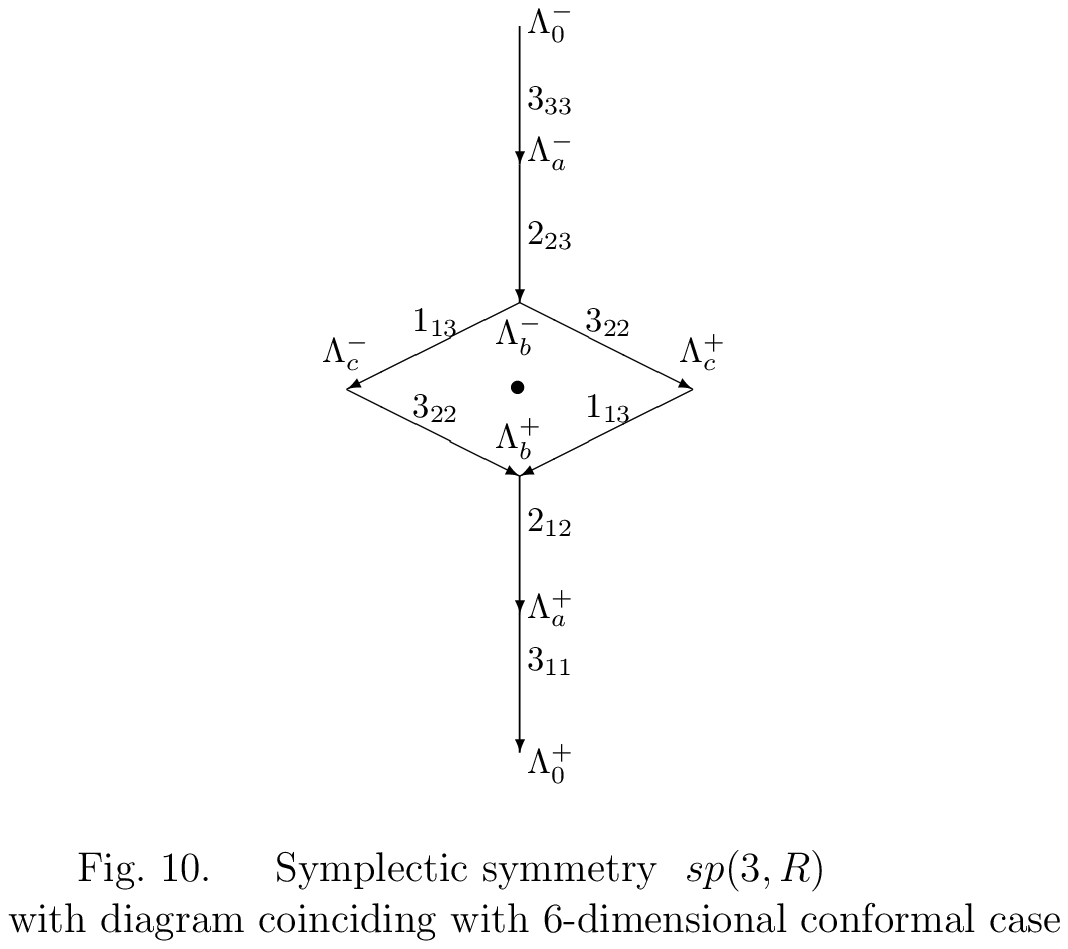}{12cm}

\fig{}{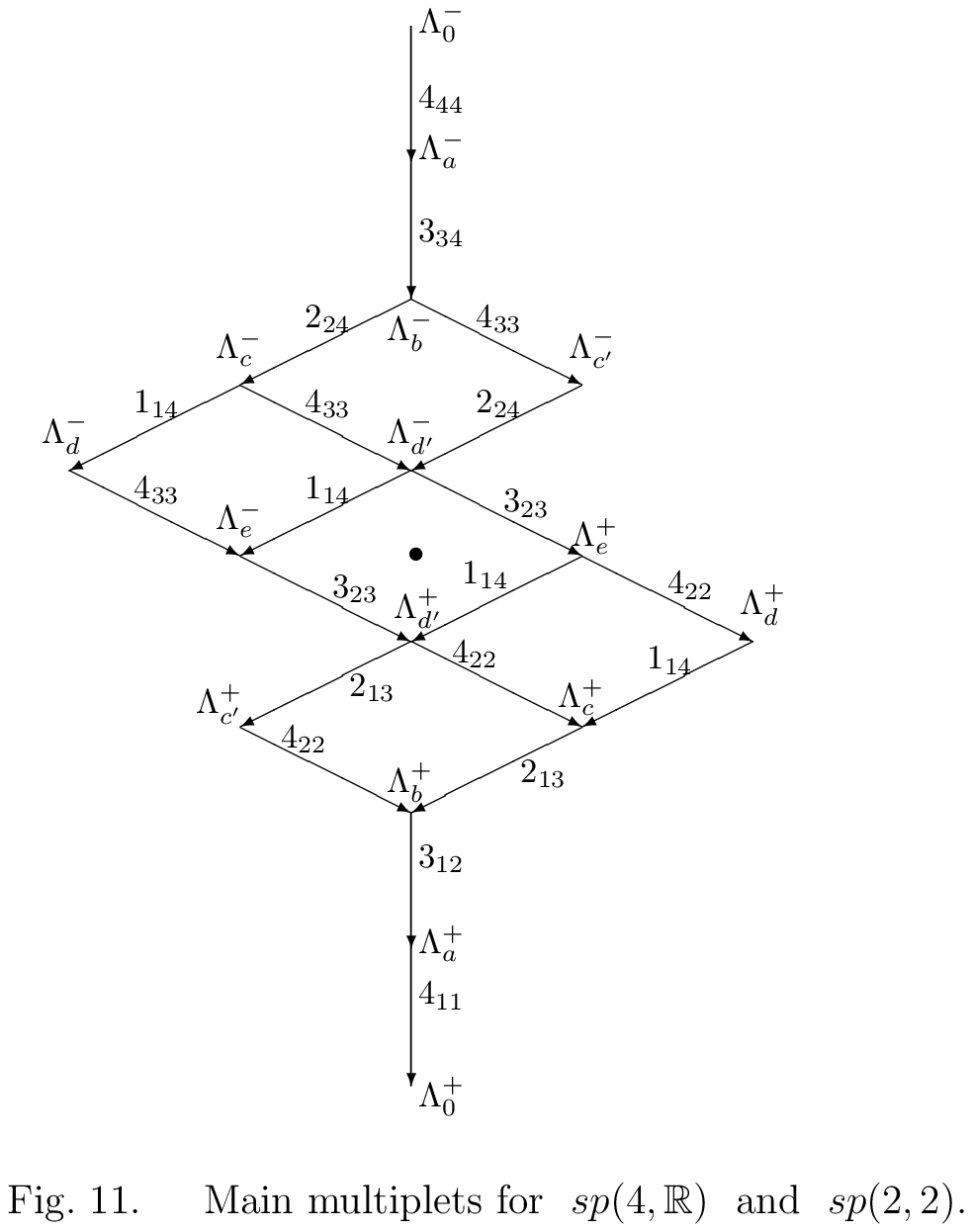}{14cm}

\fig{}{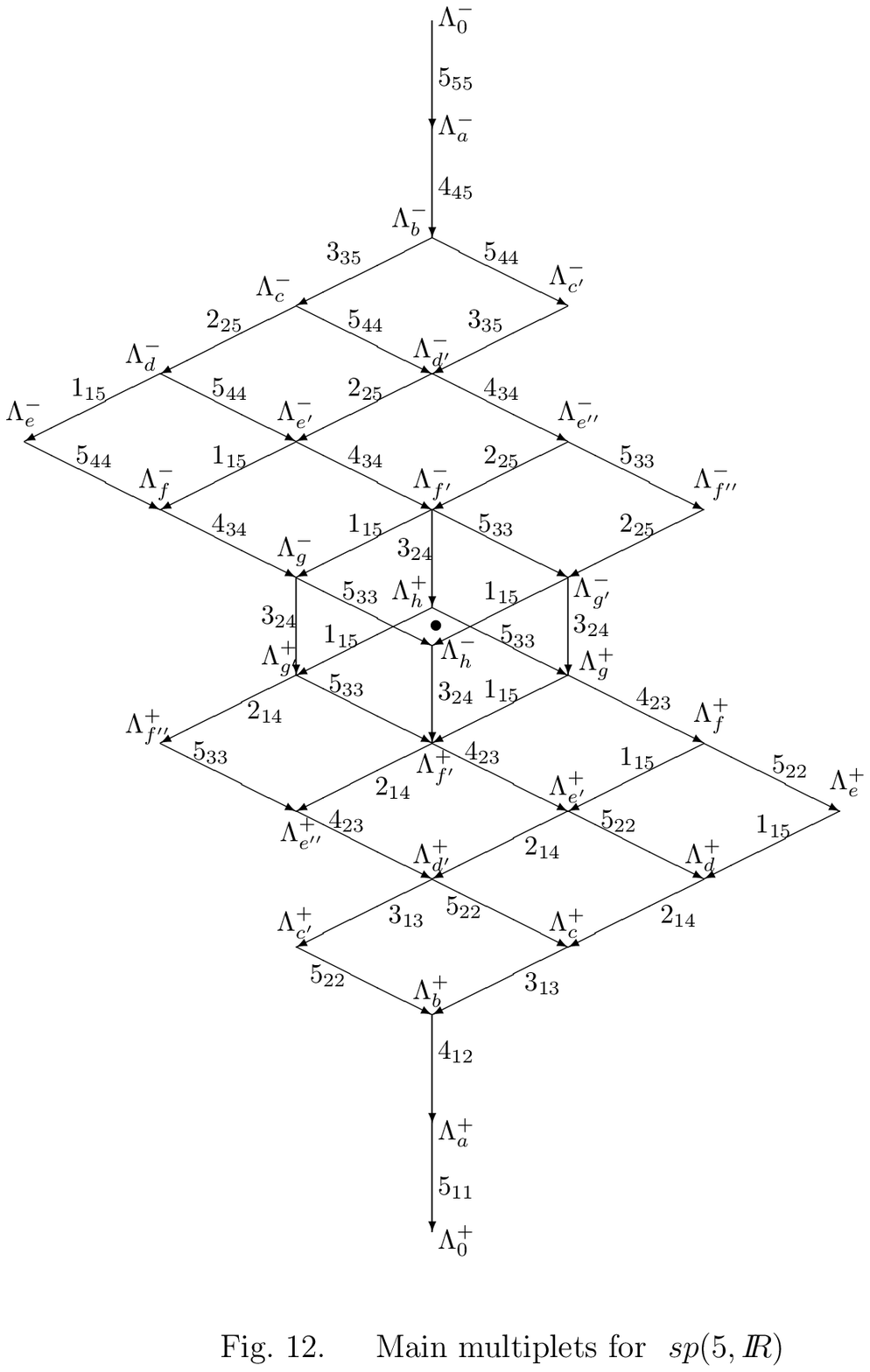}{15cm}

\fig{}{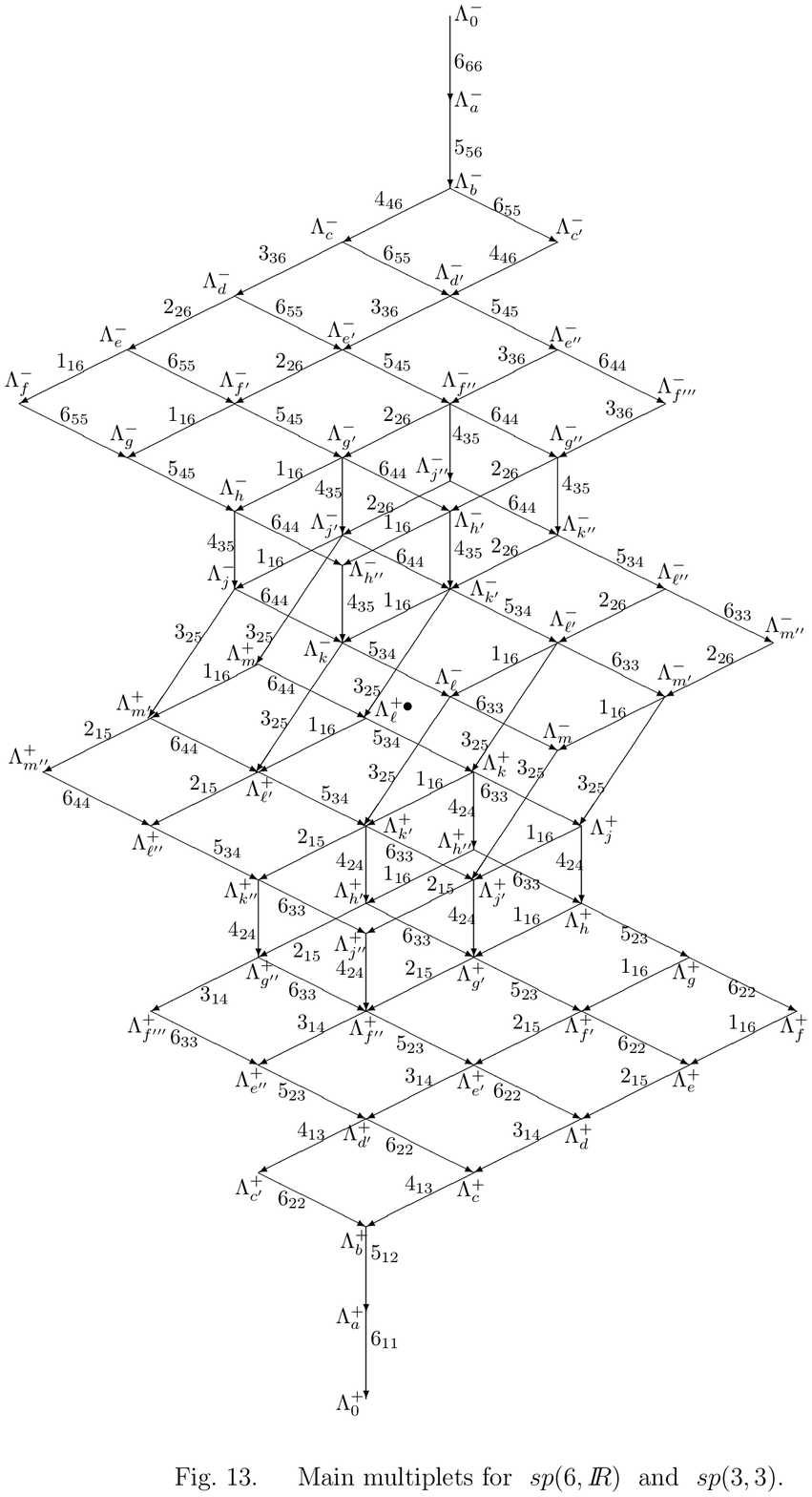}{13cm}

\section{ The  Lie algebras \blu{$E_{7(-25)}$} and
\blu{$E_{7(7)}$}}

\nt Let ~$\cg=E_{7(-25)}$. The maximal compact subgroup is ~$\ck
\cong e_6\oplus so(2)$, while ~$\cm \cong E_{6(-6)}$.

The Satake diagram \cite{Sata} is:  $$ \downcirc{{\a_1}}
\riga\black{\a_3} \riga
\black{{\a_4}}\kern-20pt\raise20pt\hbox{$\vert$}
\kern-10pt\raise40pt\hbox{$\bullet {{\scriptstyle{\a_2}}}$}
\kern-15pt\riga\black{{\a_5}} \riga\downcirc{{\a_6}} \riga\downcirc{{\a_7}}$$

The signatures of the ERs of $\cg$    are:
$$  \chi ~=~ \{\, n_1\,, \ldots,\, n_{6}\,;\, c\, \} \ ,
\qquad n_j \in \bbn\ .$$
expressed through the Dynkin labels:  \eqnn{}
&&n_i = m_i \ ,
\qquad c ~=~ -\ha (m_\ta + m_7) =  -\,\ha(
2m_1+2m_2 +   3m_3 +   \, 4m_4 + 3m_5 + 2m_6  + 2m_7)     \nn\eea
The same signatures can be used for the parabolically related exceptional Lie algebra ~$E_{7(7)}\,$~
(with ~$\cm$-factor $E_{6(6)}$).

The noncompact roots of the complex algebra ~$E_7\,$ are:   \eqnn{}
 &&\a_7 \,,~ ~\a_{17}  \,,~ ~\ldots\,,~
 \a_{67}\ , \cr &&  \a_{1,37}\,,~\a_{2,47}\,,~    \a_{17,4}  \,,~   \a_{27,4}  \,,~\cr&&
\a_{17,34}  \,,~  \a_{17,35}  \,,~ \a_{17,36}  \,,~ \a_{17,45}  \,,~ \a_{17,46}  \,,~ \cr &&
  \a_{27,45}  \,,~ \a_{27,46}   \,,~\cr&&
\a_{17,25,4}  \,,~  \a_{17,26,4}  \,,~ \a_{17,35,4}  \,,~
  \a_{17,36,4}  \,,~  \cr&&
 \a_{17,26,45}  \,,~  \a_{17,36,45}  \,,~\cr&&
       \a_{17,26,35,4}
\,,~   \a_{17,26,45,4}  \,,~ \cr&& \a_{17,16,35,4}    ~=~ \tilde{\a}\ ,  \nn\eea
given through the simple roots $\a_i$~:
\eqnn{} && \a_{ij} ~=~ \a_i + \a_{i+1} + \cdots + \a_j \ , ~i< j  \ , \cr
&& \a_{ij,k}\! =\! \a_{k,ij}\! =\! \a_i + \a_{i+1} +\cdots + \a_j +\a_k\ ,  \quad ~i< j
\ , \qquad {\rm etc.}
 \nn\eea

\nt The multiplets of the main type are in 1-to-1 correspondence
with the finite-dimensional irreps of ~$E_7\,$, i.e., they will be
labelled by  the seven positive Dynkin labels    ~$m_i\in\bbn$.

 The number of ERs in the corresponding
multiplets is equal to $$\vr W(\cg^\bac,\ch^\bac)\vr\, /\, \vr
W(\ck^\bac,\ch^\bac)\vr ~=~ 56$$
The multiplets are given in Fig.~14, \cite{Dobeseven,Dobparab}.

The Knapp-Stein operators ~$G^\pm_\chi$~ act pictorially as reflection w.r.t. the  bullet intertwining each
~$\ct_\chi^-$~ member with the corresponding ~$\ct_\chi^+$~ member.

\fig{}{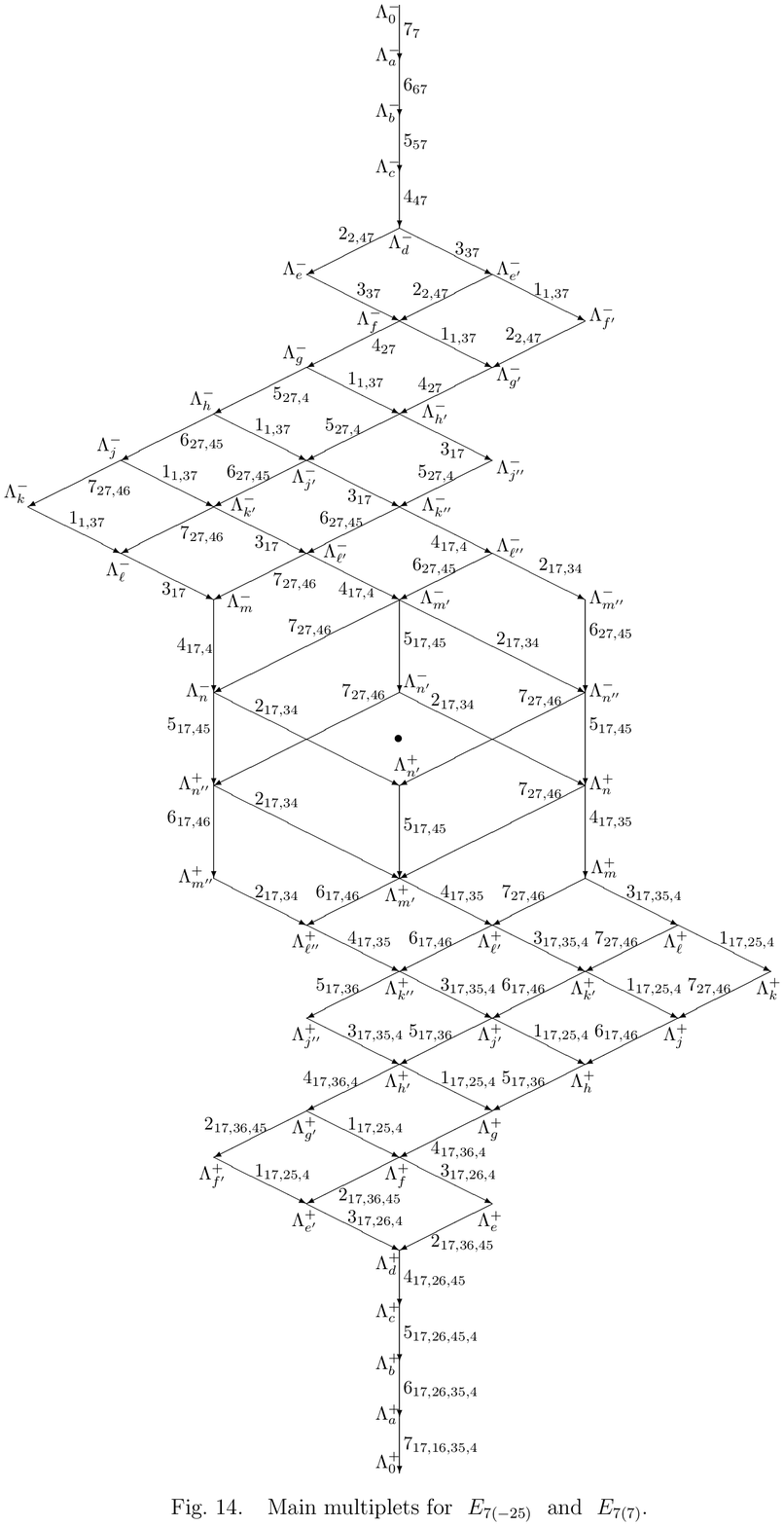}{12cm}

\section{ The  Lie algebras $E_{6(-14)}$, ~$E_{6(6)}$ and
$E_{6(2)}$}

 Let ~$\cg ~=~ E_{6(-14)}\,$.  The
maximal compact subalgebra is ~$\ck \cong so(10)\oplus so(2)$,
while ~$\cm \cong su(5,1)$.

The Satake  diagram \cite{Sata} is:
\eqn{satsixay} \underbrace{ \downcirc{{\a_1}} \riga  \black{\a_3} \riga
\black{{\a_4}}\kern-8pt\raise11pt\hbox{$\vert$}
\kern-3.5pt\raise22pt\hbox{$\circ {{\scriptstyle{\a_2}}}$}
\riga\black{{\a_5}} \riga\downcirc{{\a_6}} }\ee

The signature of the ERs of $\cg$  is:
\eqn{sgnsix} \chi = \{\, n_1\,, n_3\,,n_4\,,n_5\,,n_6\,;\, c \} \
, \quad c = d- \hel\ . \ee
expressed through the Dynkin labels as:  \be
n_i = m_i \ , \quad - c ~=~ \ha m_\ta ~=~  \ha ( m_1+2m_2 + 2m_3 +
3m_4 + 2m_5 + m_6) \ee

The same signatures can be used for the parabolically related exceptional Lie algebras ~$E_{6(6)}$ and $E_{6(2)}$ with $\cm$--factors
$sl(6,\bbr)$ and $su(3,3)$, resp.

 Further, we need the noncompact roots of the complex algebra ~$E_6\,$ :
  \eqnn{satsixzz}
&&\a_2\,,~ \a_{14}\,,~ \a_{15}\,,~\a_{16} \,,~  \a_{24}\,,~ \a_{25}\,,~\a_{26}  \\ &&
 \a_{2,4}\,,~ \a_{2,45}\,,~
 \a_{2,46}\,,~
  \a_{25,4}\,,~  \a_{15,4} \,,~\a_{26,4}  \cr
&&\a_{16,4} \,,~  \a_{15,34} \,,~ \a_{26,45} \,,~
\a_{16,34} \,,~ \a_{16,45}   \cr &&\a_{16,35}\,,~  \a_{16,35,4}\,,~
\a_{16,25,4} ~=~ \ta \  \nn
 \eea

\nt The multiplets of the main type are in 1-to-1 correspondence
with the finite-dimensional irreps of ~$\cg\,$, i.e., they will be
labelled by the six positive Dynkin labels ~$m_i\in\bbn$.

Since these algebras do not belong to the class of conformal Lie algebras (CLA) the number of ERs/GVMs in the multiplet is not given
by formula \eqref{multi}. It turns out that each such multiplet contains 70 ERs/GVMs - see Fig.~15, \cite{Dobesix,Dobparab}.
Another difference with the CLA class is that pictorially the
the Knapp-Stein operators ~$G^\pm_\chi$~ act as reflection w.r.t. the dotted line separating the
~$\ct_\chi^-$  ~ members from the ~$\ct_\chi^+$~ members (and not as reflection  w.r.t. a central dot (bullet)
as in the CLA cases).

 Note that there are five cases when the embeddings correspond to the
highest root $\ta$~: ~~$V^{\L^-} \lra V^{\L^+}$, ~$\L^+ ~=~\L^-
-m_\ta\,\ta\,$. In these five cases the weights are denoted as:
~$\L^\pm_{k''}\,$, ~$\L^\pm_{k'}\,$, ~$\L^\pm_{\tk}\,$,
~$\L^\pm_{k}\,$, ~$\L^\pm_{k^o}\,$, then:   ~$m_\ta ~=~
m_1,m_3,m_4,m_5,m_6\,$, resp.  Thus, their action coincides
with the action of the Knapp-Stein
operators ~$G^+_\chi$~ which in the  above five cases  degenerate
to differential operators as we discussed for   $so(3,2)$.

Note that the figure has the standard $E_6$ symmetry, namely,
conjugation exchanging indices $1\llra 6$, $3\llra 5$.

\fig{}{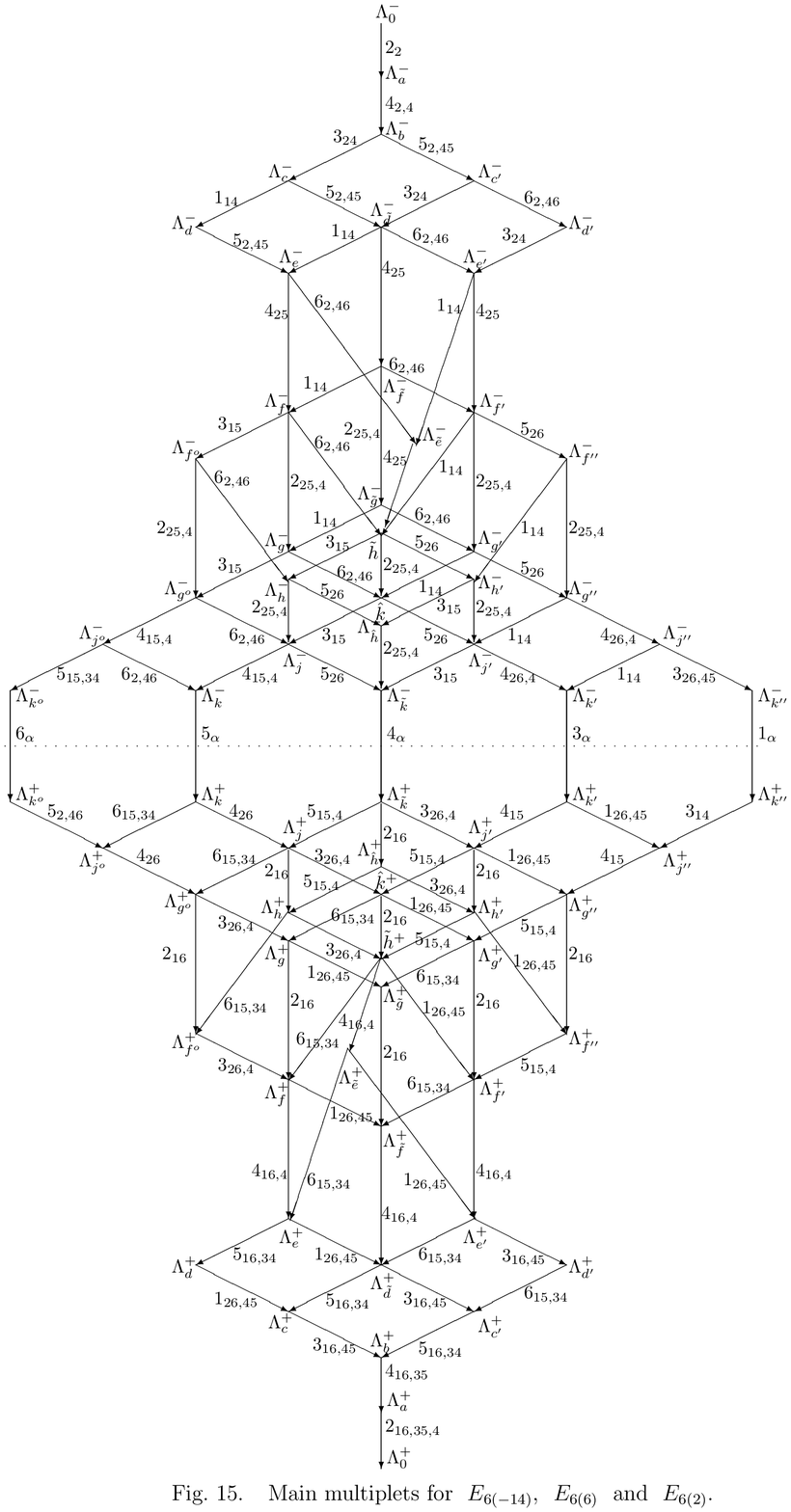}{13cm}

\section*{Acknowledgments}
It is a pleasure to thank the organizers of the VIII International Symposium "Quantum Theory and
Symmetries", and in especially Kurt Bernardo Wolf, for the hospitality. The author
has received partial support from COST action MP-1210.

\end{document}